\documentclass[12pt,preprint]{aastex}
\usepackage{amsmath}
\usepackage{amssymb}
\renewcommand{\mag}{\mbox{$\;$mag}}

\begin{document}
\title{\MakeUppercase{%
     Amplitude Fine-Structure in the Cepheid P-L Relation 
     I: Amplitude Distribution Across the RR Lyrae Instability Strip
     Mapped Using the Accessibility Restriction Imposed by the
     Horizontal Branch}}

\author{Allan Sandage}
\affil{The Observatories of the Carnegie Institution of Washington,\\
       813 Santa Barbara Street, Pasadena, CA, 91101}

\begin{abstract}
The largest amplitude light curves for both RR Lyrae (RRL) variables
and classical Cepheids with periods less than 10 days and greater than
20 days occur at the blue edge of the respective instability strips.
It is shown that the equation for the decrease in amplitude with
penetration into the strip from the blue edge, and hence the amplitude
fine structure within the strip, is the same for RRL and the Cepheids
despite their metallicity differences. However, the manifestation of
this identity is different between the two classes of variables
because the sampling of the RRL strip is restricted by the discrete
strip positions of the horizontal branch, a restriction that is absent
for the Cepheids in stellar aggregates with a variety of ages.

     To show the similarity of the strip amplitude fine structure for
RRL and Cepheids we make a grid of lines of constant amplitude in the
HR diagram of the strip using amplitude data for classical Cepheids in
the Galaxy, LMC, and SMC. The model implicit in the grid, that also
contains lines of constant period, is used to predict the correlations
between period, amplitude, and color for the two Oosterhoff RRL groups
in globular clusters. The good agreement of the predictions with the
observations using the classical Cepheid amplitude fine structure also
for the RRL shows one aspect of the unity of the pulsation processes
between the two classes of variables. 
\end{abstract}
\keywords{Cepheids --- stars: variables: other --- 
  stars: fundamental parameters --- stars: horizontal-branch}

\section{\MakeUppercase{Introduction}}
\label{sec:01}
Both RR Lyrae (RRL) variables and classical Cepheids with periods less
than 10 days and greater than 20 days have the largest light-curve
amplitudes at the blue edge of the instability strip. Because the
lines of constant period thread the strip in a sloping pattern from
high luminosity at the blue edge to lower luminosity at the red edge,
there are consequences for the Cepheid period-luminosity (PL) relation
and for the correlations for RR Lyrae stars of amplitude, period,
luminosity, and color.

     Although the variations of luminosity with amplitude (at constant
period) for the Cepheids and the correlations of period with color
and amplitude for the RRL are due to the same cause of large amplitude
at the blue edge, they manifest themselves in very different ways
because of the restriction in the luminosity of the RRL variables due
to the discrete position of the horizontal branch (HB) in the HR
diagram. No such restriction exits for the classical Cepheids in
stellar systems with a distribution of ages.

     This is the first of a projected series of three papers on the
amplitude fine-structure in the instability strip. The amplitude/P-L
relations using the observations of Cepheid variables in LMC, SMC,
and IC\,1613 is projected for Paper~II. In Paper~III the consequences
are to be set out for the existence of an amplitude bias in the
Cepheid P-L relation for determinations of the Hubble constant.

     It is appropriate to acknowledge here the goal of an earlier
attempt by \citet{Payne-Gaposchkin:59,Payne-Gaposchkin:61} and
\citet{Payne-Gaposchkin:66} to find such fine structure in the Cepheid
P-L relation using light curve shapes of Cepheids in the SMC.
Their objective was similar to ours here, but we use amplitude rather
than light curve shape. 

The plan of the paper is this.
\begin{enumerate}
   \item Details of the instability strip are set out in
   Figure~\ref{fig:01} in the next section, showing grid lines of
   constant period and amplitude within the RRL instability strip.

   \item Using Figure~\ref{fig:01}, predictions are made in
   Section~\ref{sec:03} of the correlations between amplitude and
   period (the Bailey diagram), period and color, and amplitude and
   color for RRL when the HB population-density morphology is similar
   to that in the globular cluster M3 for the 
   \citet{Oosterhoff:39,Oosterhoff:44} period group~I. 

   \item A diagram similar to Figure~\ref{fig:01} is
   Figure~\ref{fig:03} in Section~\ref{sec:04} for the period and
   amplitude grid lines using an M2/M15-like HB (Oosterhoff group~II)
   morphology. 

   \item Section~\ref{sec:05} gives predictions of RRL correlations
   using Figures~\ref{fig:01} and \ref{fig:03} where the differences
   in the RRL period-amplitude-color correlations for both Oosterhoff
   period groups I and II are compared.

   \item Section~\ref{sec:06} displays the mapping equations for the
   amplitude variation across the strip for classical Cepheids and
   RRL, showing the identity in the slope, $dA_{B}/d(B\!-\!V)$, of the
   period-color relation for both. 
\end{enumerate}

\section{\MakeUppercase{%
         Properties of the RR Lyrae Instability Strip Using an M3-like
         Horizontal Branch Morphology}}
\label{sec:02}
%
\subsection{The Discovery of the Instability Strip in the HR Diagram}
\label{sec:02:1}
The existence of an instability strip in the HR diagram for pulsating
stars was discovered by \citet{Adams:Joy:27} when they showed the
continuity and tight correlation of period and spectral type between
the long period Mira variables, the classical Cepheids, and the
cluster-type RRL stars. The finite width of the strip is displayed by
the tightness of the correlation of period and mean spectral type.

     A similar discovery, but with a less transparent discussion, was
made by \citet{Russell:27} and \citet{Shapley:27a,Shapley:27b} in
their attempt to understand the slope of the Cepheid period-luminosity
relation. They did not use the \citet{Ritter:1879} relation of
$P\rho^{1/2}=Q$, where $Q$ is nearly constant over a wide range of
luminosity and masses. The Ritter relation follows from classical
mechanics with Newton's law of inertia if the restoring force for the
radial displacement of the pulsation is gravity. 

     Using the Ritter relation, lines of constant density can be drawn
over the face of the HR diagram once masses are known. These lines
then become lines of constant period. Cepheid masses became known
once the paths of evolution in the HR diagram became known in the
1950s, not available to \citeauthor{Russell:27} or
\citeauthor{Shapley:27a} in 1927.  

     The constant period lines slant from higher luminosities toward
lower temperatures over the HR diagram. Without the temperature
restriction imposed by the existence of the instability strip, there
would be a large variation of the pulsation period at a given
luminosity, i.e., no tight Cepheid P-L relation would exit. However,
the temperature restriction discovered by \citeauthor{Adams:Joy:27}
leads to the tight period-luminosity relation.

     Either because the 1927 papers by \citeauthor{Russell:27} and
\citeauthor{Shapley:27a} were not particularly transparent, or because
interest in the Cepheid problem settled elsewhere, the Russell/Shapley
papers sank into near obscurity between 1930 and 1950. However, the
instability strip was rediscovered in the 1950s once evolution tracks
across the face of the HR diagram had been found. The language of an
``instability strip'' then emerged and became explicit. 

     The decisive development for RR Lyrae stars was made by
\citet{Schwarzschild:40} who showed that the RR Lyrae stars in the
globular cluster M3 were confined to a small, well defined, color
interval on the M3 HB. He had discovered the instability strip that
was implicit in the Adams/Joy temperature restriction. The RRL strip
is continuous with that for long period Cepheids. It is the
unification of the amplitude properties within the RRL and Cepheid
strips that we seek in this paper.

\subsection{Calibration of a Schematic Model for the RRL Instability Strip
                        That Has a Dependence of Amplitude Across the Strip}
\label{sec:02:2}
In this section we set out the methods and calibrations of a schematic
model of the RRL instability strip that has amplitude fine structure.
Those readers not interested in the details of the construction and
calibration of the model can skip to Section~\ref{sec:03} for its
application. 

     We desire an HR diagram that shows the blue and red edges for
fundamental mode pulsators, the lines of constant period, and the
lines of constant amplitude.

\subsubsection{Equations of the Red and Blue Fundamental Edges of the Instability Strip}
\label{sec:02:2:1}
We need both the slope, $dM_{V}/d(B\!-\!V)$, and the absolute
magnitude calibration of the red and blue fundamental mode edges in
the HR diagram. We take the slope from the continuity of the strip
from long period Cepheids to RRL to the dwarf Cepheids
($\delta$ Scuti stars).

     This is the sequence that was first isolated in part by
\citeauthor{Adams:Joy:27} and rediscovered through out the subsequent
literature into modern times (e.g., \citealt{Iben:67};
\citealt{Cox:74}, Figure~1;  \citealt{Cox:80}, Figure~3.1;  
\citealt{Gautschy:Saio:95}, Figure~1). 
The slope of the edges of the strip is observed to be nearly constant
over a range of 10 magnitudes from $M_{V}=-6$ to $+4$. We use this to
adopt the RRL slope to be $dM_{V}/d(B\!-\!V)=10.0$ taken from the mean
observed slopes for classical Cepheids in the Galaxy 
(\citealt{TSR:03}, hereafter \citeauthor*{TSR:03}, their Fig.~15),
and for Cepheids in the LMC and SMC 
(\citealt{STR:04,STR:09}, hereafter \citeauthor*{STR:04}, their Figure~8,
and \citeauthor*{STR:09}, their Figure~6, respectively).

     The calibration of the absolute magnitudes of these
fundamental-mode edges for the RRL have been set from the color edges
of the strip in M3, which are $(B\!-\!V)_{0}=0.27$ and 0.42 at
$M_{V}=0.52$ using an M3 reddening of $E(B\!-\!V)=0.01$. The adoption
of the mean absolute magnitude of $M_{V}=0.52$ is from the calibration
of the mean evolved HB with [Fe/H] = -1.5
\citep{Caputo:etal:00,McNamara:97,McNamara:00,Sandage:06,ST:06}
for fundamental mode variables in M3 
(\citealt{CCC:05}, hereafter \citeauthor*{CCC:05}, their Figures~1 and 5).
Therefore, the equations of the blue and red fundamental edges and the middle
ridge line in Figure~\ref{fig:01} are,
\begin{equation}
     (B\!-\!V)_{0} = -0.10 M_{V} + 0.322,                   
\label{eq:01}
\end{equation}
for the fundamental blue edge (FBE),
\begin{equation}
     (B\!-\!V)_{0} = -0.10 M_{V} + 0.472,                   
\label{eq:02}
\end{equation}
for the fundamental red edge, and
\begin{equation}
     (B\!-\!V)_{0} = -0.10 M_{V} + 0.397,                   
\label{eq:03}
\end{equation}
for the ridge line of the fundamental mode.

\subsubsection{Constructing the Lines of Constant Period}
\label{sec:02:2:2}
To map the lines of constant period we need both the slope and the
zero point in absolute magnitude. The slope is determined from the RRL
data in M3 as follows. 

     In principle, one could propose to use the vertical structure of
the horizontal branch as it is spread from the ZAHB by evolution, and
from that to trace the lines of constant period star-by-star by
comparing the periods of the brighter highly evolved stars with those
of the fainter, nearly unevolved stars, that are still near the
unevolved HB. The magnitude and color differences between such stars
with the same period would give the slope, $dM_{V}/d(B\!-\!V)$, of the
constant period lines. The data for such a procedure are set out
elsewhere \citep[][hereafter \citeauthor*{Sandage:90}]{Sandage:90}.

     However, such a star-by-star method fails because the instability
strip is not wide enough in color to produce any such pairs of
constant period variables. For example, consider the highly evolved
RRL V27 in NGC\,6981 from the photometry by \citet{Dickens:Flinn:72}
from Table~7 and Figure~11 of \citeauthor*{Sandage:90}. The star is
$0.34\mag$ brighter than the ZAHB and has the long period of 0.675 
days. We seek the color of other RRL in NGC 6981 with this period
but none exist; the strip is too narrow in color to contain them.

     However, a variation of the method exists that does not rely on
star-by-star comparisons but on ensemble averages over all stars in
the strip that are highly evolved, compared with those near the ZAHB
that are not. Consider the variation of period with color across the
strip. This period-color correlation, such as in Figures~\ref{fig:02}(c)
and \ref{fig:04}(c) later, and in Figure~5c of \citeauthor*{CCC:05}, has
dispersion about a central ridge line. Stars with the longest period
at a given color that are brighter than average have evolved from the
ZAHB. The upper envelope (longest period at a given color) in the
period-color correlation show the maximally evolved stars in the
strip. Stars on the lower period-color envelope have the shortest
period at a given color and are on the ZAHB of the cluster. 

     The upper and lower envelope lines of the period-color
correlation in M3 from the photometry by \citeauthor*{CCC:05} (their
Figures~5c) have the equations,  
\begin{equation}
     (B\!-\!V)^{\rm upper}_{0} = 0.909\log P + 0.481,       
\label{eq:04}
\end{equation}
and
\begin{equation}
     (B\!-\!V)^{\rm lower}_{0} = 0.909\log P + 0.602.       
\label{eq:05}
\end{equation}
Hence, at a given period, the envelope lines differ in color by
$0.121\mag$.

     Therefore, in an HR diagram, a line of constant period will
differ in color by $0.12\mag$ {\em at the respected edges of the strip}.
Equations~(\ref{eq:01}) and (\ref{eq:02}) for the magnitudes at the
edges of the strip at given color, when combined with
Equations~(\ref{eq:04}) and (\ref{eq:05}) give the slope of the lines
of constant period determined in this way to be
\begin{equation}
    dM_{V}/d(B\!-\!V) = 10 - 1.5/\Delta(B\!-\!V)_{\rm at\ const\ period}, 
\label{eq:06}
\end{equation}
where $\Delta(B\!-\!V)$ is the color width of the strip at constant
period. (Not to be confused with $\Delta(B\!-\!V)$ used in
Section~\ref{sec:02:2:3} for the color penetration into the strip from
the blue edge). Hence, from Equation~(\ref{eq:06}), if
$\Delta(B\!-\!V)_{{\rm const}\; P} = 0.12\mag$, then
$dM_{V}/d(B\!-\!V)=2.5$.

     Elegant as this method seems, it is sensitive to the value of the
width of the strip at constant $P$. For a change of $\pm0.01\mag$ in
$\Delta(B\!-\!V)$, the slope of the constant period lines change from
1.53 to 3.63. We estimate from the position of the extreme envelope
lines in Figure~5c of \citeauthor*{CCC:05} that the error in
$\Delta(B\!-\!V)$\footnote{%
  Note that $\Delta(B\!-\!V)$ is the maximum width given by the
  extreme envelope lines that define the strip color boundaries at the
  $3^{+}\sigma$ level given defined by Equations~(\ref{eq:04}) and
  (\ref{eq:05}). It is not the dispersion that encloses  $\pm1\sigma$
  (or $\sim\!70\%$) of the total distribution that is drawn for
  illustration in Figure~\ref{fig:02}(c).} 
for the M3 RRL is no more than $\pm0.005\mag$ giving a range of the
slope to be between 2.00 and 3.04. 

     This is steeper than the constant period slope that is observed
for the classical Cepheids in the LMC (\citeauthor*{STR:04}, Figure~9  
and Equation~(27) there), and the SMC (\citeauthor*{STR:09}, Figure~3
there) which average $dM_{V}/d(B\!-\!V)=1.6\pm0.2$. However, we cannot
expect the slope to be the same for classical Cepheids and the RRL
because the mass difference between the two classes is substantial.
Mass enters into the mean density Ritter pulsation relation of period
and mean density. 

     The absolute magnitude zero point of the lines of constant period
is fixed as follows. We again adopt the mean absolute magnitude of the
average evolved HB in M3 to be at $M_{V}=0.52$, principally from the
calibration by \citet{McNamara:97,McNamara:00} using SX~Phoenicis
variables in globular clusters as themselves are calibrated by large
trigonometric parallaxes for field members of their class.

     Because this is the mean magnitude of the average evolved HB it
is also the absolute magnitude of the midpoint line of the instability
strip defined by Equation~(\ref{eq:03}) at $(B\!-\!V)_{0}=0.345$. The
distribution of periods in M3 has an average period of
$\langle\log P\rangle=-0.25$ for type ab RRL. 
Hence, with $M_{V}=0.52$, $(B\!-\!V)_{0}=0.345$, a slope of
$dM_{V}/d(B\!-\!V)=2.5$, and $\log P=-0.25$, the equation of this line
of constant period for $\log P=-0.25$ is 
$M_{V}({\rm mean\ evolved})=2.5(B\!-\!V)_{0}-0.343$. To spread this line
across the strip for different periods, we need the mean P-L slope for
the mid ridge line. This calibration comes using the RRL-like stars in
the strip for the ``above horizontal branch'' (AHB) stars where the
P-L slope, $dM_{V}/d(\log P)$, is 2.0 \citep[][hereafter
SDT94]{SDT:94}. Hence, lines of constant period for different colors
and periods have the equation, 
\begin{equation}
   M_{V}({\rm mean\ evolved})=2.5(B\!-\!V)_{0}-2.00\log P-0.85. 
\label{eq:07}
\end{equation}

\subsubsection{Lines of Constant Amplitude Across the Strip}
\label{sec:02:2:3}
The constant amplitude lines within the strip are calculated in this
way. We make the assumption, later to be proved, that the variation of
amplitude, $A_{B}$, across the strip is the same as has been measured
for the classical Cepheids as summarized for the Galaxy by
\citeauthor*{STR:04}, Equations (30) and (34), the LMC by
\citeauthor*{STR:04} (Section~6.3 and Figure~9), and SMC by
\citeauthor*{STR:09} (Section~5 and Figure~3). These data give an
average slope of the amplitude variation with the color penetration
from the blue edge as 
\begin{equation}
   dA_{B}/d\Delta(B\!-\!V) = -8.70,                         
\label{eq:08}
\end{equation}
where $\Delta(B\!-\!V)$ is the color difference from the blue strip
border of the BFE. The negative sign means that the amplitude becomes
smaller as the $\Delta(B\!-\!V)$ color penetration from the blue edge
becomes larger. 

     The equation relating $A_{B}$ and $\Delta(B\!-\!V)$, zero-pointed
to be $A_{B}=1.8\mag$ close to the blue edge, where, by definition, 
$\Delta(B\!-\!V)=0.00$, is,
\begin{equation}
   A_{B} =  -8.70\Delta(B\!-\!V) + 1.8.                     
\label{eq:09}
\end{equation}
If the amplitude variation with color-penetration into the strip does
not depend on $M_{V}$, then the lines of constant amplitude are
parallel to the blue and red fundamental edges for all $M_{V}$
within the strip.\footnote{%
  The astute reader will note the approximation we make in
  Equation~(\ref{eq:09}) here and its consequence in
  Figures~\ref{fig:01} and \ref{fig:03} later. 
  We put $A_{B}=1.8\mag$ at the blue edge, whereas at the true blue
  and red edges, $A_{B}=0$. The approximation we make is that the rise
  in amplitude is so abrupt near the blue edge that we can put the
  maximum amplitude at the blue edge in the diagrams. The true blue
  edge, where $A_{B}=0$, is slightly to the blue of the edges drawn in
  Figs.~\ref{fig:01} and \ref{fig:03}. Same for the red edge where
  $A_{B}\sim0.4\mag$ at the drawn red edge in Figs.~\ref{fig:01} and
  \ref{fig:03}. The true red edge lies perhaps $0.02\mag$ redward of
  what is drawn in Figs.~\ref{fig:01} and \ref{fig:03}.}     

\subsection{Comment on the Lack of a Metallicity Term in Equation~(\ref{eq:07})
and in \boldmath{$M_V = f(\log P, A_{B})$} for RRL Absolute Magnitudes as
Function of Period, Color, and Amplitude} 
\label{sec:02:3}
Known since the discovery by \citet{Arp:55} and the confirmation by
\citet{Kinman:59}, Oosterhoff~II period group variables have lower
metallicity than variables in period group~I. Because Group~II RRL are
brighter than those in group~I, there is a correlation of RRL absolute
magnitude with [Fe/H] \citep[][Figures~11 and 12 for a
summary]{ST:06}. Why, then, is there no metallicity term in
Equation~(\ref{eq:07}) and in the implicit Equations~(\ref{eq:01})
with (\ref{eq:08}) for $M_{V}(P,A_{B})$ that give a calibration of the
$M_{V}({\rm RR})$ instability strip? 
We note from the grid lines in Figures~\ref{fig:01} and \ref{fig:03}
in the next sections that when either $\log P$ and $(B\!-\!V)_{0}$, or
$\log P$ and $A_{B}$ are known from observations, the $M_{V}$ absolute
magnitude can simply be read off the grid. But there is no [Fe/H]
dependence in the grid, by construction.

     The explanation is that [Fe/H] is a hidden variable that
determines the morphology of the HB, and therefor the horizontal
branch ratio (HBR, see next section), which determines the nature of
the tracks in the strip. Age zero horizontal branches in 
metal poor clusters are populated only beyond the blue edge of the
strip, and all RRL in such clusters start from tracks outside the
strip, blueward, as in Figure~\ref{fig:03} later. Higher metal
abundance clusters such as M3 have ZAHB that intersect the strip on
nearly horizontal tracks as in Figure~\ref{fig:01} below. 

     Hence, the dependence of $M_{V}$ on [Fe/H] manifests itself as a
difference in the tracks (more highly evolved and tipped in the HR
diagram as we shall see in Figure~\ref{fig:03} compared with
Figure~\ref{fig:01}). In that sense [Fe/H] is a hidden variable, not
present in the $M_{V}(P,{\rm color},A_{B})$ equations nor, by
construction, in the grid lines in Figures~\ref{fig:01} and
\ref{fig:03}. 

     This explanation is implicit in the paper by
\citet{Demarque:etal:00}, and is explicit in \citet{Bono:etal:07}
where they conclude that the metallicity effect is due more to
morphology of the HB than to a direct effect of metallicity
differences on the stellar structure of the variables, real as
that is \citep{VandenBerg:etal:00}. We could, of course, have put an
[Fe/H] dependence in the position of the ZAHB at a rate of
$dM_{V}/d$[Fe/H]$\sim0.2$, consistent with
\citeauthor{VandenBerg:etal:00}, but we see in Figure~\ref{fig:03}
later that morphology differences are the dominant effect for
Oosterhoff~II variables, although the VandenBerg effect explains most
of the $M_{V}/$[Fe/H] dependence for group~I variables, which we here
neglect because we use only the M3 tracks in Figure~\ref{fig:01}.

\subsection{Assembling the Model of the Strip} 
\label{sec:02:4}
We can now assemble the model in Figure~\ref{fig:01} using 
Equations~(\ref{eq:01})--(\ref{eq:03}) for the strip midpoint and the
edges, Equation~(\ref{eq:07}) for lines of constant period, and
Equation~(\ref{eq:09}) together with Equation~(\ref{eq:01}) for the
lines of constant amplitude. The constant period lines are shown at
intervals of $\Delta\log P=0.5\;$dex ranging from $\log P=0.00$ to
$-0.35$ (days). The lines of constant amplitude in steps of
$0.2\mag$ start with $1.8\mag$ at the blue edge and are parallel to
it. 

     A schematic horizontal branch of the M3 type is shown with width
$0.09\mag$ at a mean level at $M_{V}=0.52$. The ZAHB is put at
$M_{V}({\rm unevolved})=0.61\mag$. Three tracks of evolution, two
starting within the strip, are shown for masses of 0.68, 0.72, and
0.74 solar. These tracks paraphrase those calculated by
\citet{Dorman:92}. 

     The diagram is to be understood as only schematic, not a precise
statement of settled absolute magnitude and color for all clusters,
and only approximate for M3. It can be expected that the position of
the fundamental blue and red edges will differ from cluster to
cluster, especially if the helium composition differs, that the
amplitude penetration of Equation~(\ref{eq:09}) may also differ, and
that the lines of constant period may be slightly curved. And, of
course, that the morphology of the HB for Oosterhoff period group~II
clusters will differ (shown later in Figure~\ref{fig:03}) from cluster
to cluster. 

     However, these variations are expected to be minor compared with
the large scale schematic properties of Figure~\ref{fig:01}, which we
use  in the next section to predict the correlations between period,
amplitude and color for M3-like HB morphologies.

\section{\MakeUppercase{%
         Predicted and Observed Correlations Between Period, Amplitude,
and Color for M3-Like Horizontal Branch Morphologies}}
\label{sec:03}
The morphology of the HB drawn in Figure~\ref{fig:01} is one where the
ZAHB is populated nearly equally redward and blueward of the RRL gap.
The HB morphological ratio (HBR), introduced by \citet{Lee:90}
and used by \citet{Harris:96} and \citet{Clement:etal:01} in their
catalogs, is the blue minus red number-count divided by the total HB
population across the RRL strip. The ratio so defined as 
$(B\!-\!R)/(B+{\rm RRL}+R)$, is 0.08 for M3, meaning that there are
nearly equal numbers of stars blueward and redward of the RRL strip.

     RRL variables originate on a zero age horizontal branch and
evolve toward the AGB producing a small intrinsic width to the HB,
following tracks such as calculated early by \citet{Dorman:92}. 
The Dorman models have been summarized by \citeauthor*{SDT:94} from
which the simplified presentation of the tracks is made in
Figure~\ref{fig:01}. The mean level of the evolved HB shown in
Figure~\ref{fig:01} is put at $0.09\mag$ brighter than the ZAHB
\citep{Sandage:90,Sandage:93}. A single evolutionary track that
originates outside the strip is shown as it approaches the base of the
asymptotic giant branch \citep[e.g., Figure~13a of][]{Dorman:92}.
Three highly evolved HB stars are schematically marked by dark
triangles on it. These symbols are carried into panels (a), (c), \&
(d) of Figure~\ref{fig:02}. 

     Using Figure~\ref{fig:01} we can make predictions of the expected
correlations between period and amplitude (the Bailey diagram), period
and color, and color and amplitude for M3-like tracks.
Figure~\ref{fig:02} is a collage of such predictions.

\subsection{The Predicted Bailey Diagram of Period versus Amplitude} 
\label{sec:03:1}
Consider the predicted correlations of amplitude with period (the
Bailey diagram), made as follows. 

     The predicted amplitude at a given period along the observed mean
evolved HB is the manifold of intersections of the lines of constant
period and the $A_{B}$ amplitude at various segments of the
$M_{V}=0.52$ track in Figure~\ref{fig:01}. The prediction is shown in
Figure~\ref{fig:02}(a) as the solid line. The dashed line is the result
of reading these intersections of the period and amplitude lines in
Figure~\ref{fig:01} along the highly evolved track for mass 0.68 that
begins outside the strip. To be noted is the difference in slope of
the solid and dashed lines. This is a decisive feature of the
observations in all clusters that show highly evolved stars. The
prediction here is a success.

     Comparison with the observed correlations for M3 are shown as
Roman crosses in Figure~\ref{fig:02}(a). The observational data are from
\citeauthor*{CCC:05}, and are listed in Table~\ref{tab:01}. The
agreement with the solid-line prediction is good, although the
observed relation is slightly nonlinear. The curvature can, of course,
be produced by making Equation~(\ref{eq:09}) slightly non-linear
(curved near the maximum amplitudes), but the refinement is not made
here because it is unimportant in arguing the case.

     Figure~\ref{fig:02}(b) is the same as \ref{fig:02}(a) but with
envelope lines surrounding the ridge line drawn with $\Delta\log
P=\pm0.02\;$dex, taken from the Bailey diagram for M3 by
\citeauthor*{CCC:05}. This is not the total dispersion but is put at
$\Delta\log P=\pm0.02\;$dex so as to encompass the non-linearity of
the observed points. The envelope lines in Figure~\ref{fig:02}(b)
encompass about 70\% of the total sample whose rms in $\log P$ at
constant $A_{B}$ is larger at $\Delta\log P=\pm0.04\;$dex.

\subsection{The Color-Period and Amplitude-Color Relations} 
\label{sec:03:2}
Figure~\ref{fig:02}(c) shows the period-color prediction made by
reading Figure~\ref{fig:01} as follows.

     We select a family of constant period lines and read the color of
the intersection of each of these lines at the $M_{V}=0.52$ HB line.
The dashed locus in Figure~\ref{fig:02}(c) is the result. The
agreement with the observations listed in Table~\ref{tab:02}, shown as
Roman crosses, is good. The envelopes that encompasses 70\%
of the total sample are drawn.

     Figure~\ref{fig:02}(d) is the predicted color-amplitude relation,
made in a similar way again by reading Figure~\ref{fig:01}, following
each line of constant amplitude until it intersects the M3 HB at
$M_{V}=0.52$. We then read the color at these intersections giving the
dashed ridge line in Figure~\ref{fig:02}(d). The observations from
\citeauthor*{CCC:05}, listed in Table~\ref{tab:03}, are marked by
Roman crosses. The envelope lines for the intrinsic dispersion are
marked, predicted in an obvious way based on the vertical dispersion
between the ZAHB and the observed mean $\langle$M3$\rangle$ HB in
Figure~\ref{fig:01}. The agreement of the predicted relation and the
observations is good. 

     For an orientation with a different perspective it is useful to
recall a previous discussion \citep{Sandage:81} of the color-period
relation for M3 RRL, transformed into the temperature-period relation
for the near ZAHB HB, although used there for a different purpose.

\section{\MakeUppercase{%
         The Strip Fine Structure for the Horizontal Branch
         Morphology of Oosterhoff II Clusters}}
\label{sec:04}
The differences in the RRL correlations of period,
amplitude, and color between \citet{Oosterhoff:39,Oosterhoff:44}
period groups~I and II has a substantial literature. The ``period
shift'' phenomenon, star-by-star, rather than ensemble-average shifts
due to population-density difference along the HB, could eventually
only be explained by an absolute magnitude difference between the two
Oosterhoff groups \citep[][Figure~3]{Sandage:58}. 

     This early model for the two Oosterhoff groups has become
considerably more sophisticated since 1958, and its essence is set out
in the comparisons between Figures~\ref{fig:01} and \ref{fig:03}
later in this section. The absolute magnitude difference was
emphasized by \citet[][hereafter \citeauthor*{LDZ:90}]{LDZ:90}, as due
to evolution away from an initial ZAHB on tracks that begin blueward
of the RRL strip for group II clusters. The evolution away from the
ZAHB was used earlier to explain the vertical structure of the HB
(\citeauthor*{Sandage:90}). 

     The difference in the group~I and II tracks is shown in
Figure~\ref{fig:03}. The grid of lines of constant period and
amplitude is the same as in Figure~\ref{fig:01}, but the tracks that
start on the ZAHB outside the strip for the Oosterhoff~II clusters is
sloped in the strip. 

     Also shown in Figure~\ref{fig:03} is the HB of the anomalous,
high metallicity ([Fe/H]$=-0.5$) cluster NGC\,6441 that has an M3-like
HB morphology (HBR$\;\approx0$), yet has a large period shift relative to
Oosterhoff~I (M3-like) clusters. This requires an elevated absolute
magnitude by about $0.2\mag$, as shown in the diagram. A second
globular cluster with the same anomalous HBR morphology for its
metallicity is NGC\,6388, discovered at the same time as 
NGC\,6441 
\citep{Rich:etal:1997,Pritzl:etal:2000,Pritzl:etal:2001}.
First attempts to understand the physics of the anomaly have been made
by \citet{Sweigart:Catelan:98}, \citet{Bono:etal:97a,Bono:etal:97b},
and others on several fronts, but the problem of the physics of the
tracks appears to be still open.

     The tracks for selected masses for the M2/M15 Oosterhoff II
clusters in Figure~\ref{fig:03} show the necessary elevation in
absolute magnitude of the sloped tracks as determined observationally by 
\citet{Sandage:93},
\citet{Fernley:93},
\citet{Fernley:98a,Fernley:98b},
\citet{Carretta:etal:00},
\citet{Caputo:etal:00}
and undoubtedly others.

     The slope of the M15/M2 tracks is taken from the observations of
the horizontal branches in the color-magnitude diagram of Ml5 by
\citet{Bingham:etal:84} as merged with \citet{SKS:81}, and by
\citet{Lee:Carney:99} for M2. Two tracks for M3, both starting within
the strip on the ZAHB at $M_{V}=0.61$, are again shown at the mean
evolved luminosity of $M_{V}=0.52$.

\section{\MakeUppercase{%
       Comparisons of the RRL Correlations of Amplitude, Color,
       and Period for the Different Oosterhoff I and II HB Morphologies
       Using M3/M5 and M2/M15 as Templates}}
\label{sec:05}
The collage of four panels in Figure~\ref{fig:04} is similar to
Figure~\ref{fig:02} but with the predictions made using the different
tracks in Figure~\ref{fig:03} as templates rather than the near
horizontal tracks in Figure~\ref{fig:01}. Figure~\ref{fig:04}(a) is a
summary of the linearized period-amplitude Bailey diagrams that are
observed for M2, M3, M15, and NGC\,6441. Both the curved and the
adopted linearized observed data for M3 are shown. The star-by-star
period shifts relative to M3 are evident, as are the different slopes
for M2 and M15 compared with M3.

     Figure~\ref{fig:04}(b) shows the predicted Bailey diagrams made by
reading the intersections of the tracks with the lines of constant
period and amplitude from the grid lines in Figure~\ref{fig:03}. The
predicted period shifts relative to M3 agree well with the
observations shown in panel (a). 

     Figure~\ref{fig:04}(c) shows the observed period-color correlations
for the Oosterhoff type~I clusters M3, M5 and NGC\,6362 compared with
the Oosterhoff~II clusters of M2 and M15 and the anomalous cluster
NGC\,6441. The data for M5 are from merging the CCD photometry of 
\citet{Brocato:etal:96} with that of \citet{Storm:etal:91}, and
\citet{Caputo:etal:99}. The NGC\,6362 CCD data are from
\citet{Olech:etal:01}. The data for M2, M3, M15, and NGC\,6441 are
from the sources cited above. 

     Figure~\ref{fig:04}(d) shows the predicted period-color
correlations for M2, M3, and M15, based on the tracks and grid lines
in Figure~\ref{fig:03}. Agreement of the observed period shifts
relative to M3 in panel (c), and in the slope difference between
M2/M15 and M3/M5/6362, is excellent.

     This slope difference, so evident in the observations in panel
(c) and also in the predictions in panel (d), is, of course, due to
the near horizontal M3 track compared with the sloped M2 and M15
tracks as they cross the strip. The upward M2/M5 tracks cross the
lines of constant period at longer periods for given $(B\!-\!V)$
colors.

\section{\MakeUppercase{%
              Identity of the Amplitude Mapping Across the Strip for Cepheids
              and RRL Stars Showing Unity of the Strips}}
\label{sec:06}
We say again that the lines of constant amplitude in
Figures~\ref{fig:01} and \ref{fig:03} are from observations of
{\em classical Cepheids\/} in the Galaxy, LMC, and SMC for periods
smaller than 10 days and larger than 20 days. The slope of the
amplitude-color-penetration relation that is adopted in
Equation~(\ref{eq:08}) has been zero-pointed for the RRL by assuming
that $A_{B}=1.8\mag$ at the blue edge at $M_{V}=0.52$, and then made
parallel to the blue edge for other absolute magnitudes. Justification
of the assumption is from the excellent agreement of the observed and
predicted correlations between period, amplitude, and color in
Figures~\ref{fig:02} and \ref{fig:04}. 

     However, a more direct proof of the assumption of the unity of
the Cepheid and RRL strips as regards the amplitude variation with the
color penetration can be made by using the slope of the observed
amplitude-color correlation for RRL stars and Cepheids directly.
Figure~\ref{fig:05} shows the $A_{B}-(B\!-\!V)_{0}$ correlation of M2,
M3, M5, and M15.
The agreement of the slopes is good between the Oosterhoff~I and II
clusters. But the point to be made is that the slope here of
$dA_{B}/d(B\!-\!V)=-8.70$ for M3 is identical to the slope in
Equation~(\ref{eq:08}) for classical Cepheids despite the difference
in metallicity between the two classes of pulsators.

\section{\MakeUppercase{%
         Discussion and Summary}}
\label{sec:07}
This is the first of a three paper series where we map the instability
strip in its amplitude properties for both the RRL here and the
classical Cepheids in Papers~II and III, and where the consequent
amplitude fine structure of the Cepheid {\em period-luminosity relation\/} 
is studied.

     The mapping for RRL variables in this paper reveals properties of
the strip not available without the constraint of the RRL living on the
HB. The parameters of period, amplitude, and color are selectively
isolated by the restriction of the parameter space within the strip by
the discrete position of the HB in globular clusters, not present in
classical Cepheids. 

\begin{figure}[t] 
   \epsscale{0.90}
\plotone{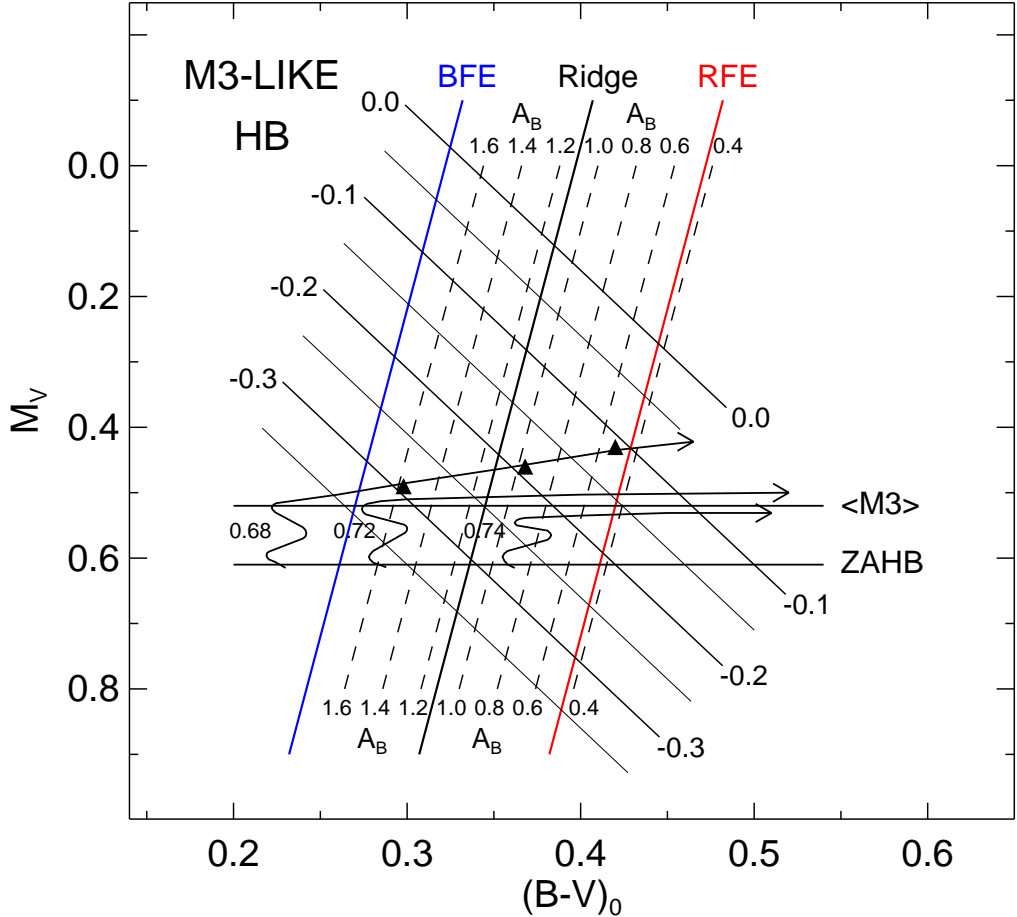}
\caption{The instability strip in the HR diagram relevant for RRL
  variables from  $0< M_{V}<+1$. The blue and red edges of the strip
  and the mid-ridge line for RRab fundamental pulsators are from
  Equations~(\ref{eq:01})--(\ref{eq:03}). The intercepts in these
  equations are set by requiring  $(B\!-\!V)_{0}=0.27$ and 0.42 at the
  strip edges for $M_{V}({\rm mean\ evolved})=0.52$, based on CCD data
  for M3 (\citeauthor*{CCC:05}). The lines of constant period are put
  with a slope of $dM_{V}/d(B\!-\!V)=2.5$. A schematic age zero HB for
  an M3-like HB morphology is put at $M_{V}({\rm ZAHB})=0.61$. Masses
  are marked next to the tracks, based on \citet*{Dorman:92} models.
  Three, more highly evolved, stars are shown schematically on these
  tracks as dark triangles that are repeated in later diagrams.}
  \label{fig:01}
\end{figure}

\begin{figure}[t] 
   \epsscale{0.90}
\plotone{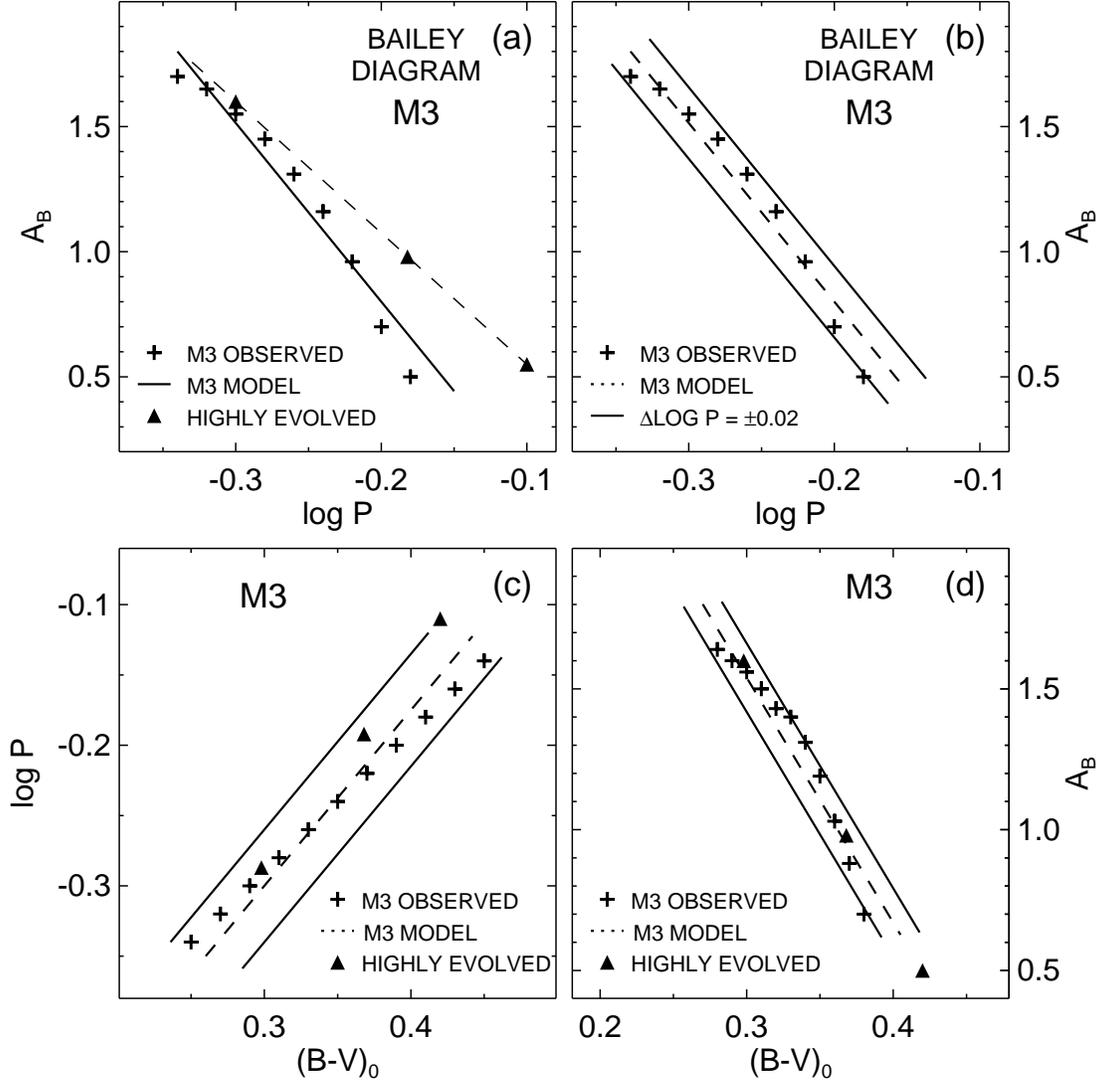}
\caption{A collage of correlations for stars on the HB and
  evolutionary tracks in Figure~\ref{fig:01}. Panel (a) is the
  predicted and observed period-amplitude Bailey diagram. The solid
  line is the prediction by reading Figure~\ref{fig:01} as described
  in the text. The Roman crosses show the ridge-line M3 observations
  from \citeauthor*{CCC:05}, listed in Table~\ref{tab:01}.
  Panel (b) is the same as (a) but with the observed envelope lines
  with $\Delta\log P=\pm0.02\;$dex put around the central ridge-line
  prediction. Panel (c) is the predicted and observed M3 period-color
  relation implicit in Figure~\ref{fig:01}. The crosses are the
  observations from \citeauthor*{CCC:05}, listed in Table~\ref{tab:02}.
  Panel (d) shows the color-amplitude prediction from
  Figure~\ref{fig:01} compared with the \citeauthor*{CCC:05} data
  listed in Table~\ref{tab:03}. The envelope lines in panels (b), (c),
  and (d) are from the observations put at about the $\pm1.5\sigma$
  level.} 
  \label{fig:02}
\end{figure}

\begin{figure}[t] 
   \epsscale{0.90}
\plotone{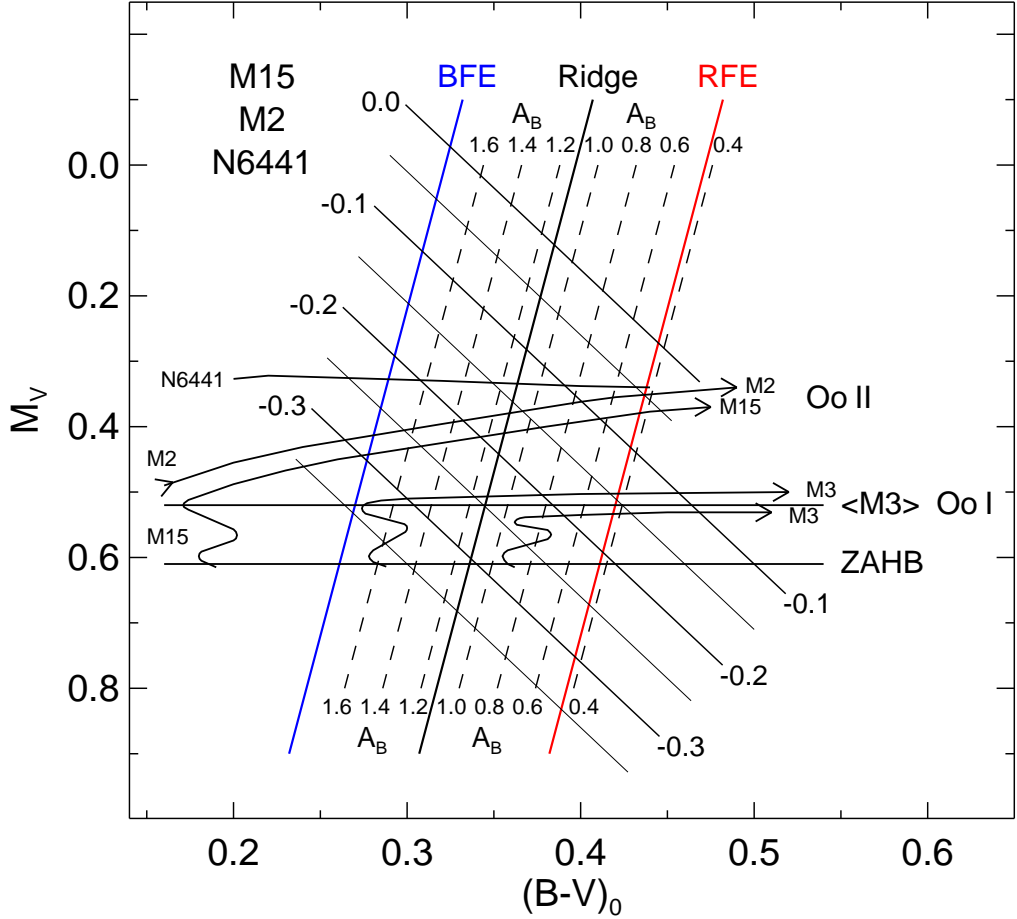}
\caption{Grid lines of constant period and amplitude from
  Figure~\ref{fig:01}. The nearly horizontal HB of M3-like tracks
  starting from an ZAHB inside the strip are contrasted with the
  sloping tracks of the Oosterhoff~II period group with M2 and M15 as
  templates, with both tracks starting from an ZAHB that is outside
  the strip. The slight variation of the position of the ZAHB for
  different metallicities over the range of [Fe/H] between $-1.5$ and
  $-2.2$ \citep[][Figure~9]{VandenBerg:etal:00,ST:06} is neglected.}
  \label{fig:03}
\end{figure}

\begin{figure}[t] 
   \epsscale{0.90}
\plotone{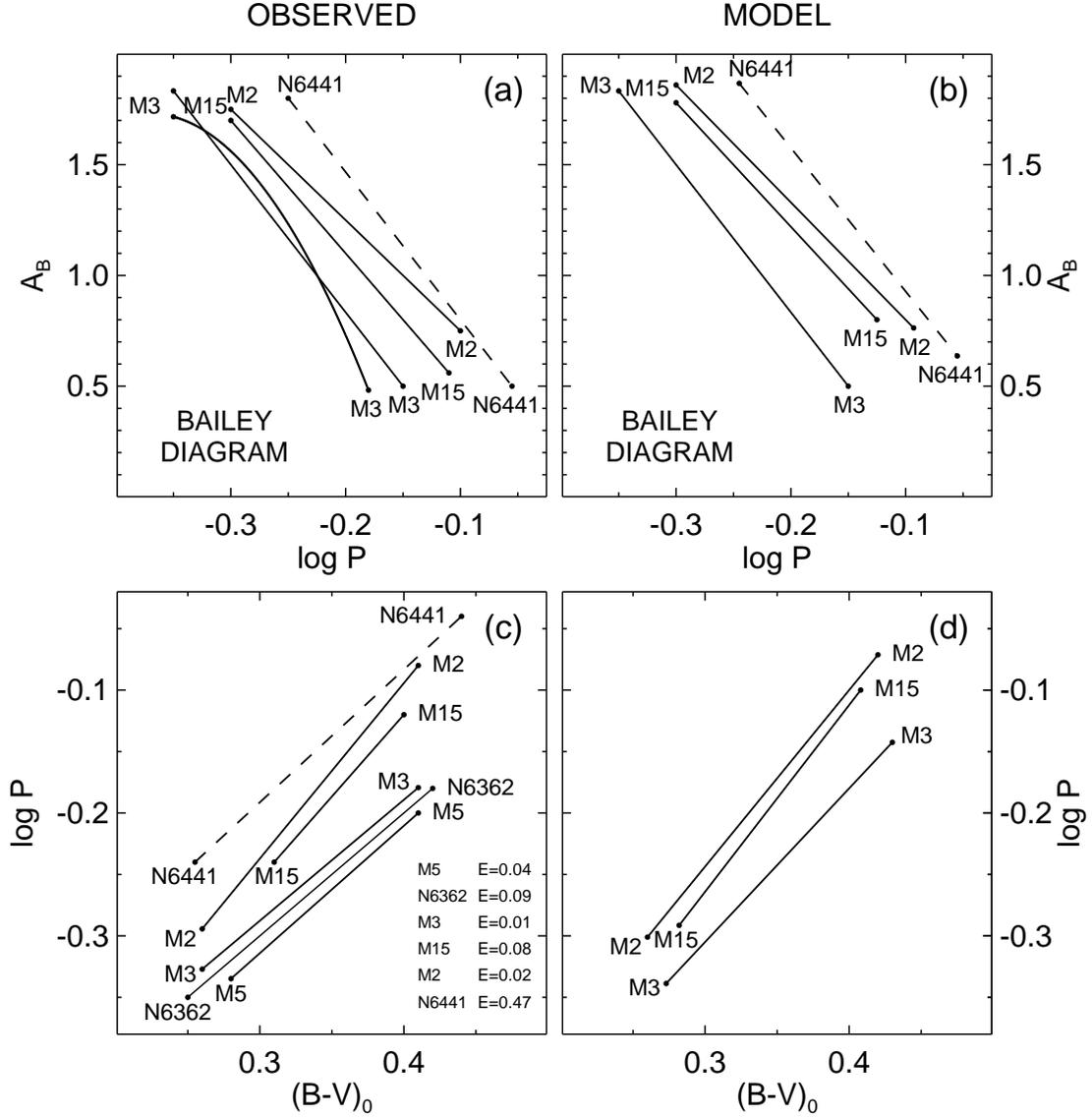}
\caption{Panel (a); the observed linearized period-amplitude Bailey
  diagrams for the four globular clusters M3 (Oosterhoff period group
  I), M2 and M15 (Oosterhoff period group II), and the anomalous
  second parameter cluster NGC\,6441. The predicted $P$-$A_{B}$ relation for
  M3, taken from Figure~\ref{fig:02} is shown with the linearized M3
  relation based on the observations of \citeauthor*{CCC:05}; (b)
  Predicted $A_{B}$-$P$ relations using Figure~\ref{fig:03} for the same
  four clusters; (c) the observed $P$-$(B\!-\!V)_{0}$ correlation for
  six globular clusters showing the difference in the slopes between
  Oosterhoff I and II clusters; the adopted reddenings are listed; (d)
  predicted $P$-color correlation for M3, M2, and M15 based on
  Figure~\ref{fig:03}.}
  \label{fig:04}
\end{figure}

\begin{figure}[t] 
   \epsscale{0.90}
\plotone{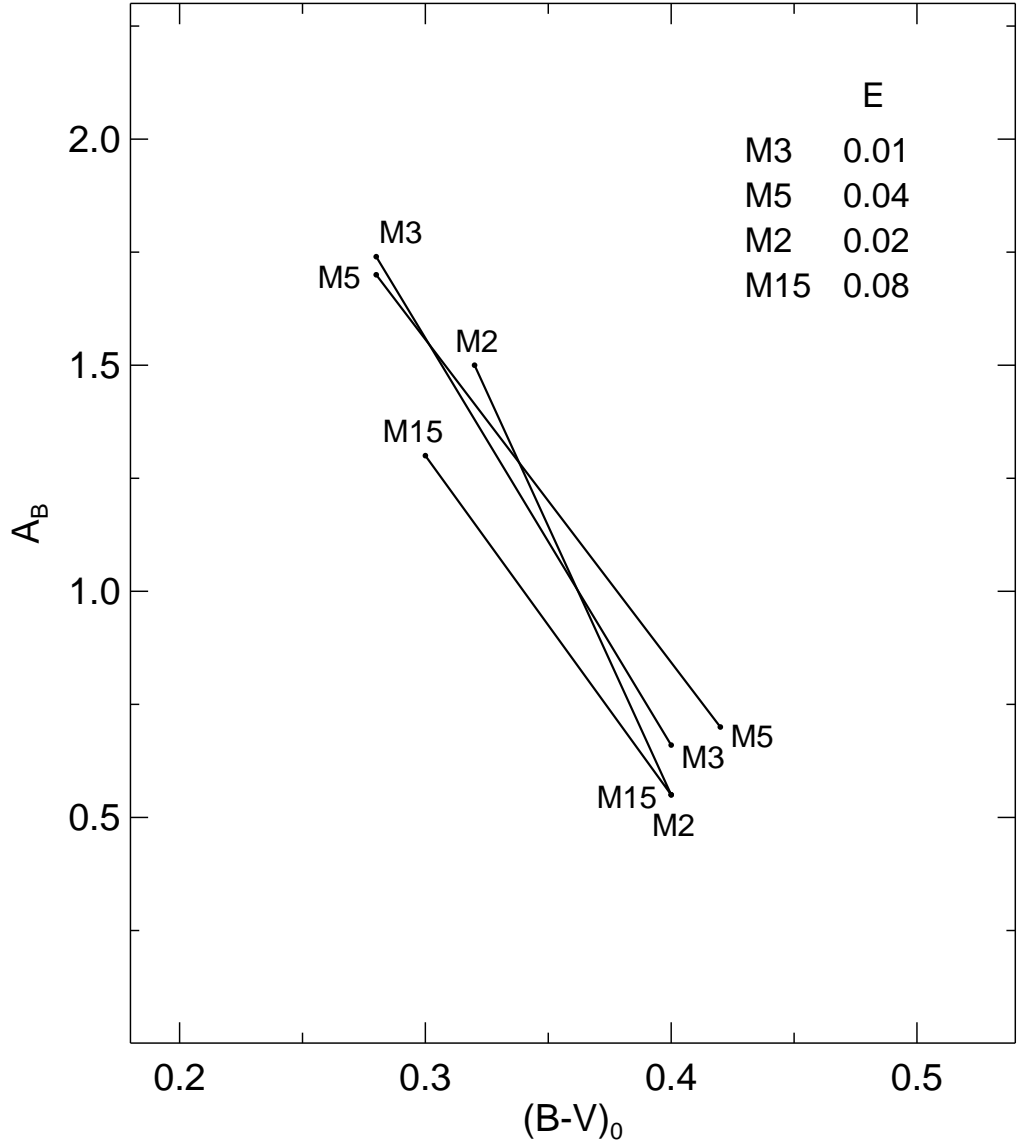}
\caption{Demonstration that the slope of the color-amplitude relation,
  corrected for reddening, is also the same for Oosterhoff period I
  and II clusters. The slope, $dA_{B}/d(B\!-\!V)$, for M3 at $-8.70$
  is also the same as for classical Cepheids from
  Equation~(\ref{eq:08}).} 
  \label{fig:05}
\end{figure}

The conclusions are these:
\begin{enumerate}
   \item The identity of amplitude variation between the Cepheids and
   the RRL is proved by adopting the observed amplitude variation
   within the strip for Cepheids in the Galaxy, LMC, and SMC to model
   the variation in the RRL variables, and to show, thereby, excellent
   agreement between the predicted and the observed correlations of
   period, amplitude, and color for the RRL stars. What appears to be
   so different between the Cepheids and the RRL in the correlations
   of parameters is shown to be only different manifestations of an
   underlying unity, explained by the model in Figures~\ref{fig:01}
   and \ref{fig:03} caused by the confinement of the RRL to the HB.

   \item The positions of the blue and red strip borders for
   fundamental mode pulsators in the RRL domain are in
   Equations~(\ref{eq:01})--(\ref{eq:03}), taken from the HR strip
   positions of Cepheids in the Galaxy, LMC, and SMC
   (\citeauthor*{STR:04,STR:09}) and scaled to the absolute magnitudes
   of the RRL. The lines of constant period are from
   Equation~(\ref{eq:07}). The adopted slope of the variation of the
   amplitude for various color penetrations, $\Delta(B\!-\!V)$, into
   the strip from the blue side is in Equation~(\ref{eq:08}),
   zero-pointed at $A_{B}=1.8$ at the fundamental blue edge in
   Equation~(\ref{eq:09}). 

   \item The predictions, read from Figures~\ref{fig:01} and
   \ref{fig:03}, for the $P$-$A_{B}$ Bailey diagrams, the $P$-color, and
   the $A_{B}$-color correlations for Oosterhoff period groups~I and
   II clusters are in Figures~\ref{fig:02} and \ref{fig:04}, and
   compared there with the observations. The excellent agreement
   between the observations and the schematic predictions in the
   absolute zero points in these diagrams is a proof that the model
   in Figures~\ref{fig:01} and \ref{fig:03} has merit. 

   \item Many facts of the observed $P$/$A$-color correlations for the
   manifold of cluster variables of both RRL Oosterhoff period groups
   are reproduced in Figures~\ref{fig:02} and \ref{fig:04}. They are
   shown to be explained by the different HB morphologies of the
   evolution tracks within the instability strip that greatly restrict
   accessibility to only those parts of the strip that are occupied by
   the HB. Paramount is the difference in the $dM_{V}/d(B\!-\!V)$
   slope of the M2 and M15 tracks in Figure~\ref{fig:03} compared to
   that for the group I cluster M3.

   \item Proof that the slope of the amplitude variation with color
   across the strip is the same for Cepheids and RRL is in
   Figure~\ref{fig:05}, where the $dA_{B}/d(B\!-\!V)$ slope for the
   RRL (M3 in particular) is the same as for the Cepheids. This
   completes the proof of the unity of the amplitude properties of the
   strips in both. 

   \item The tight correlations of period and color, corrected for
   reddening, in Figure~\ref{fig:04}(c) and its prediction in
   \ref{fig:04}(d), and the equally tight correlation between $A_{B}$
   and $(B\!-\!V)_{0}$ in Figure~\ref{fig:05} provide two new methods to
   measure the $E(B\!-\!V)$ reddening for other clusters from their
   RRL variables relative to the adopted reddenings of M2, M3, M5, and
   M15 with an accuracy of $\pm0.02\mag$ estimated from the scatter in
   these diagrams. 

   \item The results of this paper form the preliminaries for Paper~II
   that will address the fine structure in the Cepheid period-luminosity
   relation that depends on amplitude. 
\end{enumerate}

\acknowledgments
It is a pleasure to thank, in the order that they entered the problem,
Garry Kim, M.~D., 
Babak Tashakkor, M.~D., 
Michael Lin, M.~D., 
Joshua Ellenhorn, M.~D., and 
Stephen Koehler, M.~D., 
each of whom made parts of this paper possible. 
Thanks also are for Bernd Reindl for making the diagrams and the text
ready for submission, G.~A. Tammann and A.~Gautschy for reading and
informally refereeing an early draft, John Grula, librarian and
chief editorial officer of the Carnegie Observatories for his
liaison with the press in seeing the paper through the editorial
process, and to the Carnegie Institution for post retirement
research facilities.



\clearpage

\begin{deluxetable}{cccc}
\tablewidth{0pt}
\tabletypesize{\footnotesize}
\tablecaption{\sc Observed Ridge-Line Period-Amplitude
Bailey Diagram for M3 From Data By \citeauthor*{CCC:05}\label{tab:01}}
\tablehead{
 \colhead{$\log P$}      & 
 \colhead{$A_{B}$}       & 
 \colhead{$\log P$}      & 
 \colhead{$A_{B}$}            
}
\startdata
  -0.34 & 1.70 & -0.24 & 1.16 \\
  -0.32 & 1.65 & -0.22 & 0.96 \\
  -0.30 & 1.55 & -0.20 & 0.70 \\
  -0.28 & 1.45 & -0.18 & 0.50 \\
  -0.26 & 1.31 & -0.16 & 0.20 \\
\enddata
\end{deluxetable}

\begin{deluxetable}{cccc}
\tablewidth{0pt}
\tabletypesize{\footnotesize}
\tablecaption{\sc Observed Ridge-Line Period-Color
Relation for M3 From Data by \citeauthor*{CCC:05}\label{tab:02}}
\tablehead{
\multicolumn{4}{c}{$E(B\!-\!V)=0.01$} \\
 \colhead{$\log P$}        & 
 \colhead{$(B\!-\!V)_{0}$} &            
 \colhead{$\log P$}        & 
 \colhead{$(B\!-\!V)_{0}$}
}
\startdata
  -0.36 & 0.23 & -0.24 & 0.35 \\
  -0.34 & 0.25 & -0.22 & 0.37 \\
  -0.32 & 0.27 & -0.20 & 0.39 \\
  -0.30 & 0.29 & -0.18 & 0.41 \\
  -0.28 & 0.31 & -0.16 & 0.43 \\
  -0.26 & 0.33 & -0.14 & 0.45 \\
\enddata
\end{deluxetable}

\begin{deluxetable}{cccc}
\tablewidth{0pt}
\tabletypesize{\footnotesize}
\tablecaption{\sc Observed Ridge-Line Color-Amplitude
Relation for M3 From Data by \citeauthor*{CCC:05}\label{tab:03}}
\tablehead{
\multicolumn{4}{c}{$E(B\!-\!V) =0.01$} \\
 \colhead{$(B\!-\!V)_{0}$} & 
 \colhead{$A_{B}$}         &              
 \colhead{$(B\!-\!V)_{0}$} & 
 \colhead{$A_{B}$}                      
}
\startdata
  0.28 & 1.64 & 0.34 & 1.31 \\
  0.29 & 1.60 & 0.35 & 1.19 \\
  0.30 & 1.56 & 0.36 & 1.03 \\
  0.31 & 1.50 & 0.37 & 0.88 \\
  0.32 & 1.43 & 0.38 & 0.70 \\
  0.33 & 1.40 & 0.39 & 0.40 \\
\enddata
\end{deluxetable}

\clearpage

\end{document}